\documentclass[12pt,a4paper]{conference}

\usepackage{fancyhdr}
\usepackage{graphicx,amsmath,amssymb,cite,epsfig}
\usepackage{multind}
\makeindex{author} \makeindex{subject}

\newcommand{\beq}{\begin{equation}}
\newcommand{\eeq}[1]{\label{#1}\end{equation}}
\newcommand{\eeqn}{\end{equation}}

\newcommand{\beqa}{\begin{eqnarray}}
\newcommand{\eeqa}[1]{\label{#1}\end{eqnarray}}
\newcommand{\eeqan}{\end{eqnarray}}

\let\bar=\overbar

\newcommand{\Dslash}{\not{\hbox{\kern-4pt $D$}}}
\newcommand{\dslash}{\not{\hbox{\kern-2pt $\del$}}}

\newcommand{\msb}{{\bar{\ssstyle M \kern -1pt S}}}

\begin{document}

\Chapter{Dynamical Coupled-Channels Model Analysis of $\pi$-N
Scattering and Electromagnetic Pion Production Reactions}
           {Dynamical Coupled-Channels Model Analysis \dots}{B. Juli\'a-D\'\i az}
\vspace{-6 cm}\includegraphics[width=6 cm]{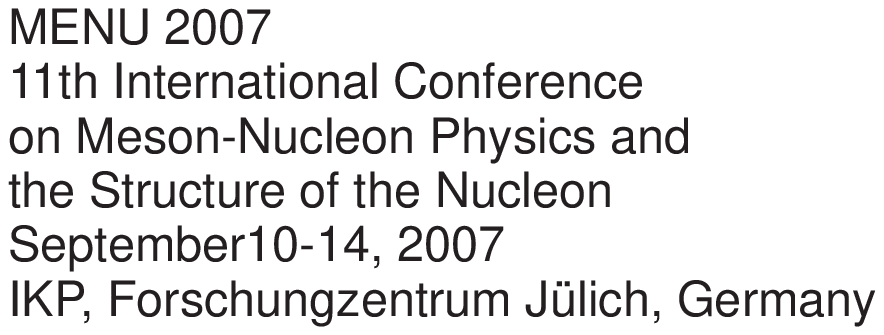}
\vspace{4 cm}

\addcontentsline{toc}{chapter}{{\it N. Author}} \label{authorStart}

\begin{raggedright}

{\it B. Juli\'a-D\'\i az}\index{author}{Julia-Diaz, B.}\\

Departament de Estructura i Constituents de la Materia,\\
University of Barcelona,\\
Spain, 08028\\
\bigskip\bigskip

\end{raggedright}

\begin{center}
\textbf{Abstract}
\end{center}

The ability of the coupled-channels model (MSL) developed in
recently in Ref.~\cite{msl} to account simultaneously for the $\pi
N$ scattering data and the $\pi$ photoproduction reactions on the
nucleon is presented. An accurate description of $\pi N$ scattering
has been obtained. A preliminary description of $\pi$
photoproduction is also discussed.

\section{Introduction}

Understanding the intricate dynamics leading to baryon resonance
production and subsequent decay is essential to deepen our
understanding on the confinement mechanisms of QCD. It is now well
recognized that a coupled-channels approach is needed to extract the
nucleon resonance $(N^*)$ parameters from the data of $\pi N$ and
electromagnetic meson production reactions. With the recent
experimental developments~\cite{lee-reviewa,lee-reviewb}, such a
theoretical effort is needed to analyze the very extensive data from
Jefferson Laboratory (JLab), Mainz, Bonn, GRAAL, and Spring-8. To
cope with this challenge, a dynamical coupled-channels model (MSL)
for meson-baryon reactions in the nucleon resonance region has been
developed recently~\cite{msl}.

The details of the MSL model are given in Ref.~\cite{msl} and will not
be discussed here. Similar to the earlier
works using meson-exchange models of pion-nucleon
scattering, see Ref.~\cite{lee-reviewa} for a review, the starting point of
the MSL model is a set of Lagrangians
describing the interactions between mesons (including the photon)
($M$ =$\gamma$, $\pi, \eta$ , $\rho, \omega$, $\sigma, \dots$) and
baryons ($B = N, \Delta, N^*, \dots$). By applying a unitary transformation
method~\cite{sl-1,sko}, an effective Hamiltonian is then derived from
the considered Lagrangian. All of these interactions are
{\it energy independent}, an important feature of the MSL formulation.

Once the Hamiltonian has been obtained, the coupled-channels
equations for $\pi N$ and $\gamma N$ reactions are derived by using
the standard projection operator technique~\cite{feshbach}, as given
explicitly in Ref.~\cite{msl}. The obtained scattering equations
satisfy the two-body ($\pi N, \eta N$, $\gamma N$) and three-body
($\pi\pi N$) unitarity conditions. The $\pi\Delta$, $\rho N$ and
$\sigma N$ resonant components of the $\pi \pi N$ continuum are
generated dynamically.

The constructed model can describe up to now very well almost all of the
empirical $\pi N$ amplitudes in $S$, $P$, $D$, and $F$ partial waves of
SAID~\cite{said}. We will also show that the predicted differential cross
sections and target polarization asymmetry are in good agreement with
the original data of elastic $\pi^{\pm} p \rightarrow \pi^{\pm} p$ and
charge-exchange $\pi^- p \rightarrow \pi^0 n$ processes.

\section{Dynamical coupled-channels equations}

\begin{figure}[t]
\vspace{20pt}
\centering
\includegraphics[width=10cm,angle=-0]{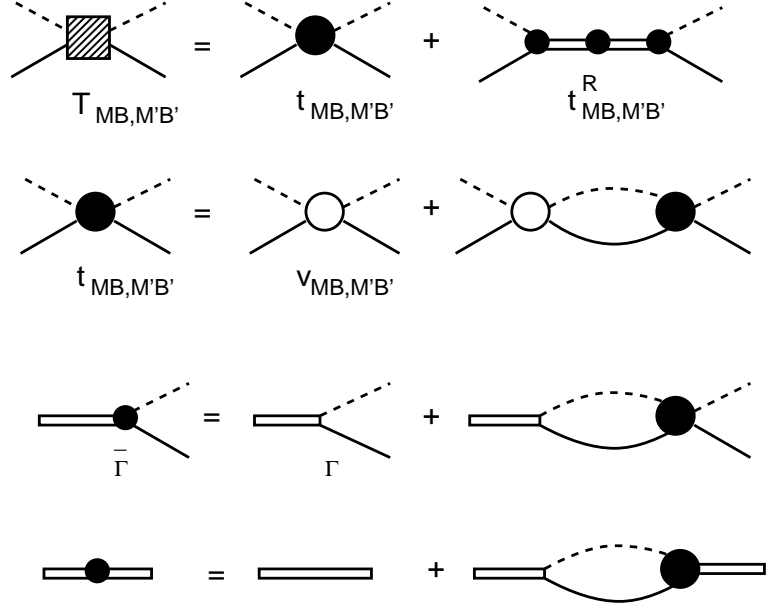}
\caption{Graphical representation of Eqs.(\ref{eq:tmbmb})-(\ref{eq:nstar-mb}).}
\label{fig:tmatmbmb}
\end{figure}

The meson-baryon ($MB$) scattering equations derived in Ref.~\cite{msl}
are illustrated in Fig.~\ref{fig:tmatmbmb}. Explicitly, they are defined
by the following equations
\begin{eqnarray}
 T_{\alpha,\beta}(E)  &=&
 t_{\alpha,\beta}(E)
+
 t^R_{\alpha,\beta}(E) \,,
\label{eq:tmbmb}
\end{eqnarray}
where $\alpha, \beta = \gamma N, \pi N, \eta N, \pi\pi N$. The full
amplitudes, e.g. $T_{\pi N,\pi N}(E)$, $T_{\eta N,\pi N}(E)$,
$T_{\pi N,\gamma  N}(E)$ can be directly used to calculate $\pi N$,
$\pi N \to \eta N$ and $\gamma N \to \pi N$ scattering observables.
The non-resonant amplitude $t_{\alpha,\beta}(E)$ in
Eq.~(\ref{eq:tmbmb}) is defined by the coupled-channels equations,
\begin{eqnarray}
t_{\alpha,\beta}(E)= V_{\alpha,\beta}(E)
+\sum_{\delta}
V_{\alpha,\delta}(E) \;
G_{\delta}(E)    \;
t_{\delta,\beta}(E)  \,
\label{eq:nr-tmbmb}
\end{eqnarray}
with
\begin{eqnarray}
V_{\alpha,\beta}(E)= v_{\alpha,\beta}
+Z^{(E)}_{\alpha,\beta}(E)\,.
\label{eq:veff-mbmb}
\end{eqnarray}

The second term in the right-hand-side of Eq.~(\ref{eq:tmbmb}) is the resonant
term defined by
\begin{eqnarray}
t^R_{\alpha,\beta}(E)= \sum_{N^*_i, N^*_j}
\bar{\Gamma}_{\alpha \rightarrow N^*_i}(E) [D(E)]_{i,j}
\bar{\Gamma}_{N^*_j \rightarrow \beta}(E) \,,
\label{eq:tmbmb-r}
\end{eqnarray}
with
\begin{eqnarray}
[D^{-1}(E)]_{i,j} = (E - M^0_{N^*_i})\delta_{i,j} - \bar{\Sigma}_{i,j}(E)\,,
\label{eq:nstar-g}
\end{eqnarray}
where $M_{N^*}^0$ is the bare mass of the resonant state $N^*$, and the
self-energies are
\begin{eqnarray}
\bar{\Sigma}_{i,j}(E)= \sum_{\delta}\Gamma_{N^*_i\rightarrow \delta} G_{\delta}(E)
\bar{\Gamma}_{\delta \rightarrow N^*_j}(E) \,.
\label{eq:nstar-sigma}
\end{eqnarray}
The dressed vertex interactions in Eq.~(\ref{eq:tmbmb-r}) and
Eq.~(\ref{eq:nstar-sigma}) are (defining
$\Gamma_{\alpha\rightarrow N^*}=\Gamma^\dagger_{N^* \rightarrow \alpha}$)
\begin{eqnarray}
\bar{\Gamma}_{\alpha \rightarrow N^*}(E)  &=&
{ \Gamma_{\alpha \rightarrow N^*}} + \sum_{\delta}
t_{\alpha,\delta}(E)
G_{\delta}(E)
\Gamma_{\delta \rightarrow N^*}\,,
\label{eq:mb-nstar} \\
\bar{\Gamma}_{N^* \rightarrow \alpha}(E)
 &=&  \Gamma_{N^* \rightarrow \alpha} +
\sum_{\delta} \Gamma_{N^*\rightarrow \delta}
G_{\delta }(E)t_{\delta,\alpha}(E) \,.
\label{eq:nstar-mb}
\end{eqnarray}

\section{Meson-baryon interaction}
\label{sec:cal}

\begin{figure}[t]
\vspace{25pt}
\begin{center}
\mbox{\epsfig{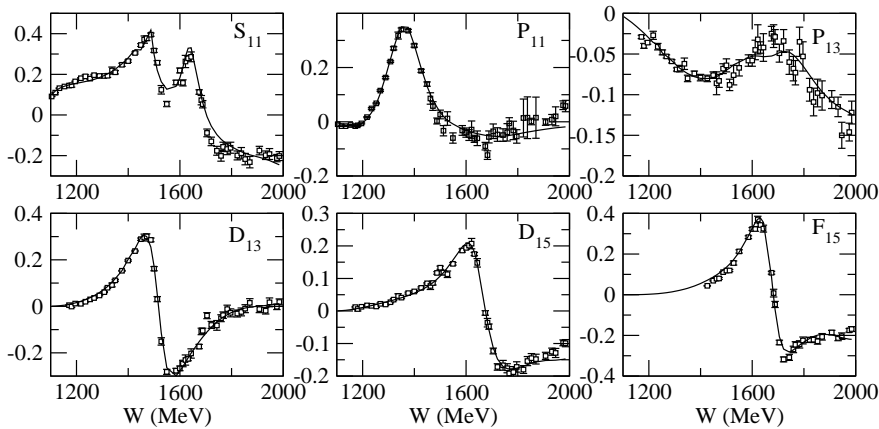}}
\end{center}
\caption{Real part of $T_{\pi N, \pi N}$ for some
isospin $1/2$ partial waves compared to the SAID
energy independent extraction.
\label{faa}}
\end{figure}

We proceed in the following way. First we consider the meson-baryon
interactions involving $\pi N, \eta N, (\pi \Delta, \sigma N, \rho N)$
and use the extensive database for $\pi N \to \pi N$ (and also
the $\pi N \to \eta N$) to fix the non-resonant parameters entering in
the phenomenological lagrangians. Once the meson-baryon is fixed we will,
in a first stage, leave it unchanged and produce a first description of
the single meson photoproduction data. In the next step a combined analysis
will need to be performed.

To solve the coupled-channels integral equations,
Eq.~(\ref{eq:nr-tmbmb}), without introducing any further
simplification we need to regularize the matrix elements of
$v_{MB,M'B'}$. We include at each meson-baryon-baryon vertex a form
factor of the following form \beq F(\vec{k},\Lambda)=\left[{
|\vec{k}|^2 \over |\vec{k}|^2+\Lambda^2}\right]^2 \eeq with
$\vec{k}$ being the meson momentum. For the meson-meson-meson the
same form is used with $\vec{k}$ being the momentum of the exchanged
meson.

With the non-resonant amplitudes generated from solving Eq.~(\ref{eq:nr-tmbmb}),
the resonant amplitude $t^R_{MB,M'B'}$ Eq.~(\ref{eq:tmbmb-r}) then depends
on the bare mass $M^0_{N^*}$ and the bare $N^*\rightarrow MB$ vertex
functions. It is worth recalling that the resonance amplitude will necessarily
contain information about the non-resonant piece, as is apparent from
Eq.~(\ref{eq:tmbmb-r}). As discussed in Ref.~\cite{msl}, these bare $N^*$
parameters can perhaps be taken from a hadron structure calculation which
{\it does not} include coupling with meson-baryon continuum states or
meson-exchange quark interactions. Unfortunately, such information is not
available to us. We thus use a parameterization given explicitly in
Ref.~\cite{jlms07}.

\begin{figure}[t]
\vspace{25pt}
\begin{center}
\mbox{\epsfig{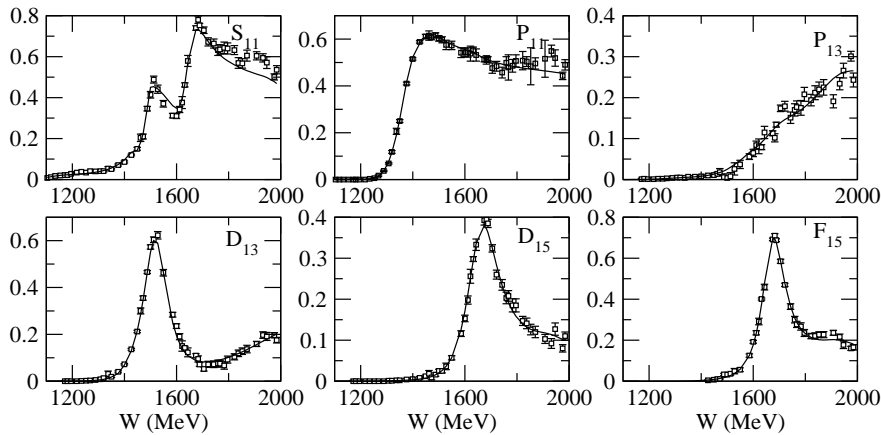}}
\end{center}
\caption{Imaginary part of $T_{\pi N, \pi N}$ for some
isospin $1/2$ partial waves compared to the SAID
energy independent extraction.
\label{fab}}
\end{figure}

\begin{figure}[tbp]
\vspace{20pt}
\begin{center}
\mbox{\epsfig{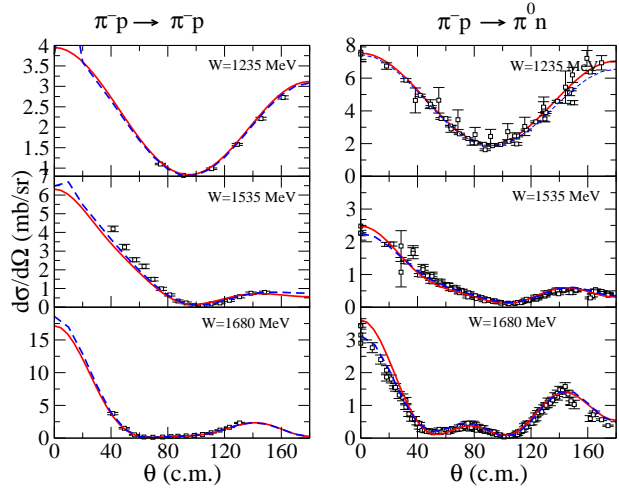}}
\end{center}
\caption{ Differential cross section for several
different center of mass energies. Solid red curve corresponds
to our model while blue dashed lines correspond to the SP06
solution of SAID~\cite{said}. All data have been obtained
through the SAID online applications.~Ref.~\cite{saiddb}.
\label{f12}}
\end{figure}

\begin{figure}[tbp]
\vspace{20pt}
\begin{center}
\mbox{\epsfig{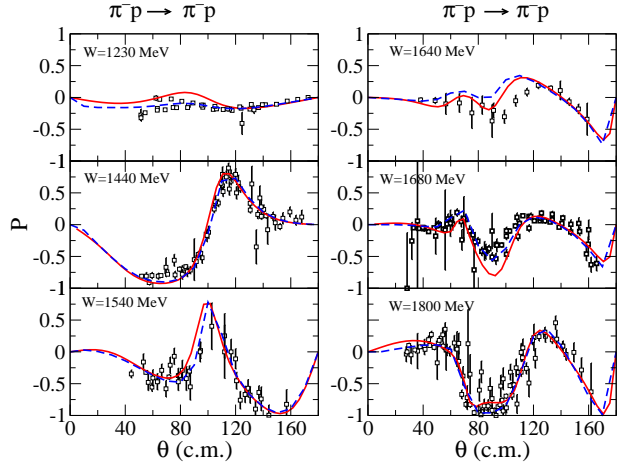}}
\end{center}
\caption{Target polarization asymmetry, $P$, for several
different center of mass energies. Solid red curve corresponds
to our model while blue dashed lines correspond to the SP06
solution of SAID~\cite{said}. All data have been obtained
through the SAID online applications.~Ref.~\cite{saiddb}.
\label{f13}}
\end{figure}

In figures~\ref{faa} and~\ref{fab} we depict the real and imaginary
part of $T_{\pi N, \pi N}$ matrix compared to the energy independent
extraction of the GWU group~\cite{said}. A comparison with experimental
data for differential cross section and target polarization asymmetry
is given in figures~\ref{f12} and~\ref{f13}.

Our model is further checked by examining our predictions of
the total cross sections $\sigma^{tot}$ which can be calculated
from the forward elastic scattering amplitudes by using the
optical theorem.

\begin{figure}[th]
\vspace{25pt}
\begin{center}
\mbox{\epsfig{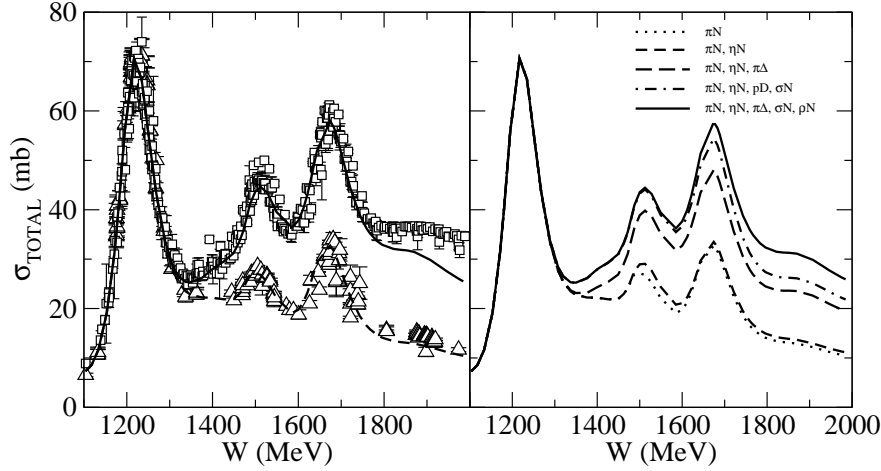}}
\end{center}
\caption{Left: The predicted total cross sections of the $\pi^- p\to X$
(solid curve) and $\pi^-p \to \pi^- p + \pi^0 n$ (dashed curve) reactions
are compared with the data. Open squares are the  data on $\pi^-p\to X$
from Ref.~\cite{pdg}, open triangles are obtained by adding the
$\pi^-p \to \pi^-p$ and $\pi^-p\to \pi^0n$ data obtained from Ref.~\cite{pdg}
and SAID database~\cite{saiddb} respectively. Right: Show how the predicted
contributions from each channel are added up to the predicted total cross
sections of the $\pi^- p\to X$.
\label{pim}}
\end{figure}

The predicted $\sigma^{tot}$ (solid curves) along with the resulting
total elastic scattering cross sections $\sigma^{el}$ compared with
the data of $\pi^- p$ reaction are shown in Fig.~\ref{pim}. Clearly,
the model can account for the data very well within the experimental
errors. Equally good agreement with the data for $\pi^+ p$ reaction
is achieved. In the right side of Fig.~\ref{pim}, we show how the
contributions from each channel add up to get the total cross sections.

The contributions from $\pi \Delta$, $\rho N$ and $\sigma N$
intermediate states to the $\pi^- p \rightarrow \pi\pi N$ total
cross sections calculated from our model can be seen in the right side of
Fig.~\ref{pim}. These predictions remain to be verified by the future
experiments. The existing $\pi N \rightarrow \pi\pi N$ data are not
sufficient for extracting $model$ $independently$ the contributions from
each unstable channel.

As mentioned above, the fit to $\pi N$ elastic scattering can not
determine well the bare $N^* \rightarrow \pi \Delta, \rho N, \sigma N$
parameters. Thus the results for these unstable particle channels
must be refined by fitting the $\pi N \rightarrow \pi\pi N$ data, this
is currently being pursued~\cite{jklms07}.

\section{Photoproduction reactions}

\begin{figure}[tb]
\vspace{25pt}
\begin{center}
\mbox{\epsfig{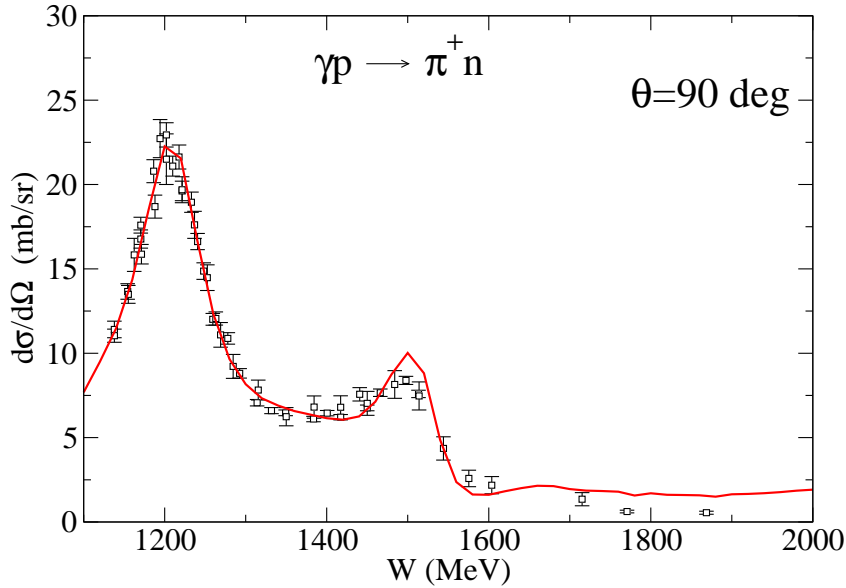}}
\end{center}
\caption{Differential cross section for the reaction
$\gamma p \to \pi^+n$ at $\theta=90$ deg as function
of the center of mass energy, W.
\label{dcspi}}
\end{figure}

With the hadronic parameters determined in the previous fit to
meson-baryon experimental data we proceed to analyze the extensive
data base of $\pi$ photoproduction. Here the only parameters that
need to be determined are the bare $\gamma N \to N^*$ vertex
interactions.

The strategy is to start with the bare helicity amplitudes
of resonances at the values given by the PDG~\cite{pdg}. Then, we
allow small variations with respect to those values and also
in a preliminary step small variations of a selected set of
non-resonant parameters. At the present stage we can only
present preliminary results which are at the present
time being further improved and will be reported elsewhere.

\begin{figure}[h]
\vspace{25pt}
\begin{center}
\mbox{\epsfig{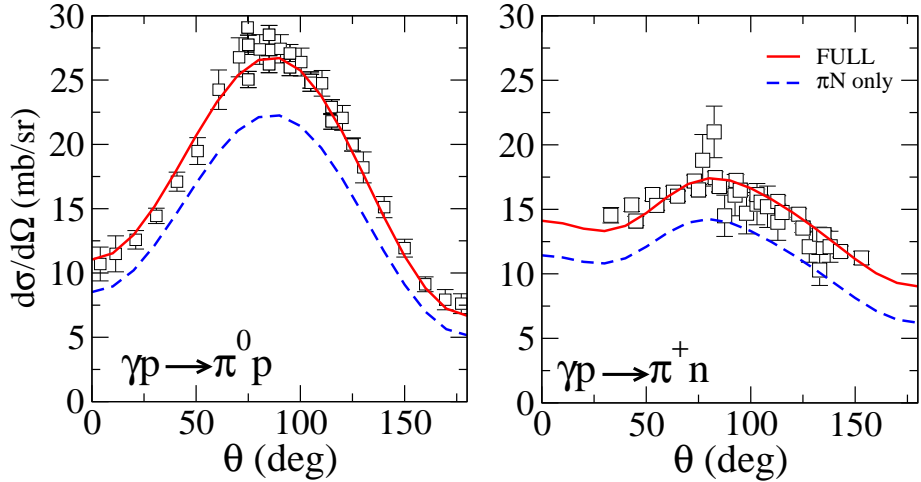}}
\end{center}
\caption{Differential cross section in the $\Delta$ region. The
full line corresponds to the current full model, the dashed line
only considers intermediate $\pi N$ intermediate states in the
photoproduction process.
\label{dcs_efb}}
\end{figure}

In figure~\ref{dcspi} we present a comparison of the current
model and the experimental differential cross section data for
the reaction $\gamma p \to \pi^+n $ at a fixed angle, $\theta=90$
(deg). First, our main emphasis is put on understanding the region
up to $1.6$ GeV extending in that way previous works where
only the $\Delta$ (1232) region was studied~\cite{sl-1,sl-2,jlss07}.
In figure~\ref{dcs_efb} we depict angular distributions for both
$\pi^+ n$ and $\pi^0 p$ photoproduction in the $\Delta(1232)$ region.
The effect of intermediate meson-baryon states different from $\pi N$
is also depict. The importance of multi-step processes is clear and
confirms previous studies done in similar frameworks.

\section{Future Developments}

The model described in detail in Refs.~\cite{msl,jlms07} has already
been used to study $\pi N$ scattering and $\pi$ photoproduction reactions
as presented in this contribution. Being the main and most important
interest of these studies the extraction and interpretation of
baryon resonances by analyzing the extant photo and electro production
experimental data we are now beginning to perform simultaneous studies
of meson-baryon, single meson electro (photo)production~\cite{jlmss07}
and two-meson photoproduction.

At the same time an important effort is being pursued to reliably
extract meaningful resonance parameters from the coupled-channels
formalism~\cite{suzukikun}.

The simultaneous consideration of other meson-baryon channels,
such as $\omega N$~\cite{mark} or kaon-hyperon channels is being
pursued within the same framework.

\section*{Acknowledgments}
It is a pleasure to thank T.-S. H. Lee, A. Matsuyama and T. Sato who
collaborated in everything presented here. I want to thank also the
hospitality of the theory group at JLAB where part of this work was done.
This work is partially supported by Grant No. FIS2005-03142 from MEC (Spain)
and FEDER and European Hadron Physics Project RII3-CT-2004-506078. The
computations were performed at NERSC (LBNL) and Barcelona Supercomputing
Center (BSC/CNS) (Spain).


\end{document}